# Theoretical Investigation of Optical Intersubband Transitions and Infrared Photodetection in $\beta$-(Al$_x$Ga$_{1-x}$)$_2$O$_3$/Ga$_2$O$_3$ Quantum Well Structures

Joseph E. Lyman[1, a)] and Sriram Krishnamoorthy[1, b)]

[1] The University of Utah, Salt Lake City, Utah 84112-9202, U.S.A.
[a)] **Author to whom correspondence should be addressed:** *joseph.e.lyman@utah.edu*
[b)] *sriram.krishnamoorthy@utah.edu*

*Abstract*

We provide theoretical consideration of intersubband transitions designed in the ultra-wide bandgap Aluminum Gallium Oxide ((Al$_x$Ga$_{1-x}$)$_2$O$_3$)/Gallium Oxide (Ga$_2$O$_3$) quantum well system. Conventional material systems have matured into successful intersubband device applications such as large area quantum well infrared photodetector (QWIP) focal plane arrays for reproducible imaging systems but are fundamentally limited via maximum conduction band offsets to mid- and long-wavelength infrared applications. Short- and near-infrared devices are technologically important to optical communications systems and biomedical imaging applications, but are difficult to realize in intersubband designs for this reason. In this work, we use a first-principles approach to estimate the expansive design space of monoclinic $\beta$-(Al$_x$Ga$_{1-x}$)$_2$O$_3$/Ga$_2$O$_3$ material system, which reaches from short-wavelength infrared (1–3 $\mu$m) to far infrared (>30 $\mu$m) transition wavelengths. We estimate the performance metrics of two QWIPs operating in the long- and short-wavelength regimes, including an estimation of high room-temperature detectivity (~ 10$^{11}$ Jones) at the optical communication wavelength $\lambda_p = 1.55$ $\mu$m. Our findings demonstrate the potential of the rapidly maturing (Al$_x$Ga$_{1-x}$)$_2$O$_3$/Ga$_2$O$_3$ material system to open the door for intersubband device applications.

## I. Introduction

Infrared devices are utilized in a diverse set of applications. Night vision technology operates in the near infrared (0.78–1 $\mu$m), optical communications systems in the short-wavelength infrared (1–3 $\mu$m), and thermal imaging devices work in the mid- (3–6 $\mu$m) and long- (6–15 $\mu$m) wavelength infrared regions [1]. The very long- (15–30 $\mu$m) and far- (30–100 $\mu$m) infrared is important in biomedical sensing and astronomical observation and includes the ever-elusive terahertz spectrum. Optical intersubband transitions in quantum well structures offer a tunable range of emission and detection wavelengths attractive for several applications. Many quantum well-based devices have revolutionized electronics and optoelectronics namely the resonant tunneling diode (RTD) [2-5], quantum cascade laser (QCL) [6-8], and quantum well infrared photodetector (QWIP) [9-15].

The development of highly uniform, large-area epitaxial growth methods such as molecular beam epitaxy (MBE) and metal-organic vapor-phase epitaxy (MOVPE) enables large-scale fabrication of RTDs, QCLs, and QWIPs. The remarkable development of large-area QWIP focal plane arrays (FPAs) by NASA's Goddard Space Flight Center demonstrates the technological advantage offered by this low-cost, highly reproducible imaging device for applications such as long-wavelength FPAs for the NASA Landsat surveillance missions [16, 17]. Conventional GaAs-based devices are fundamentally limited to operation wavelengths not much shorter than ~ 6 $\mu$m due to a maximum usable direct gap conduction band offsets (CBOs) of around 0.34 eV [18]. Short-wavelength and near-infrared QWIPs have been challenging to realize, purely limited by the material system's maximum conduction band offset (CBO). Exploring optical transitions in novel material systems provides opportunity for expanding the design space of such devices.

Large CBOs are available in wide bandgap materials , which has driven research into materials such as II-VI [19-24] and AlGaN/GaN [25-30] material systems. The II-VI system is based on complex quaternary alloys, which complicate design and material synthesis. It has been demonstrated that the internal polarization of conventional c-plane GaN materials is detrimental to the wavelength-tunability of optical intersubband transitions [25]. Nonpolar GaN eliminates these internal fields, but is very challenging to grow and wafer size is limited, whereas such aspects are necessary for a scalable device.

Gallium Oxide ($Ga_2O_3$) is an ultra-wide bandgap transparent conducting oxide; its unique properties have motivated research into device applications such as high-power transistors [31] and solar-blind, deep-ultra-violet photodetectors [32]. The most stable phase, $\beta$-$Ga_2O_3$, has a bandgap of 4.6 eV and monoclinic crystal structure [33]. A large CBO is achieved in $\beta$-$Ga_2O_3$ by alloying with Aluminum. $\beta$-$(Al_xGa_{1-x})_2O_3$ is predicted to be stable at concentrations up to $x = 0.7$ [34]; at this concentration, the bandgap theoretically increases to 6.3 eV with a type-II band-alignment with nearly no valence band discontinuity such that CBOs of up to 1.7 eV are theoretically available [33, 34]. These band discontinuity calculations have been experimentally verified in $\beta$-$(Al_xGa_{1-x})_2O_3$/$Ga_2O_3$ heterostructures for compositions up to $x = 0.2$ [35, 36]. In addition, recent density functional theory (DFT) predictions show gamma-valley electrons isolated by at least 2 eV to neighboring valleys [37]. This is important for ISB devices because inter-valley scattering via satellite valleys distracts from intersubband transitions and is detrimental to device performance.

With the increasing availability of large-area $\beta$-$Ga_2O_3$ bulk single-crystal substrates suitable for epitaxial growth [38, 39], this emerging semiconductor material holds promise for device applications. The successful demonstration of high-quality epitaxial thin-films of $\beta$-$Ga_2O_3$ using metal-organic vapor phase epitaxy (MOVPE) [40-43] and molecular-beam epitaxy (MBE) [44-48] techniques motivates research into heterostructure-based devices. With a large CBO and rapid materials development, the $\beta$-$(Al_xGa_{1-x})_2O_3$/$Ga_2O_3$ material system clearly shows potential for scalable ISB devices operating in shorter wavelengths than conventional materials.

In the following work, we use a self-consistent Schrodinger-Poisson solver to predict the ISB design space of single β-$(Al_xGa_{1-x})_2O_3/Ga_2O_3$ quantum well structures which spans peak transition wavelengths as large as 80 $\mu$m and as short as 2 $\mu$m. Using first-principles models and approximations, we estimate performance metrics (dark current, responsivity, and detectivity) of two β-$(Al_xGa_{1-x})_2O_3/Ga_2O_3$ QWIPs operating at detection wavelengths of $\lambda_p = 1.55$ $\mu$m (SWIR) and $\lambda_p = 9$ $\mu$m (LWIR). Our results demonstrate long-wavelength performance on par with conventional cryogenically cooled detectors and short-wavelength performance comparable to existing devices. Section II introduces the simulation methods and physical models used to estimate spectral response and performance characteristics. Section III discusses the intersubband design space of $(Al_xGa_{1-x})_2O_3/Ga_2O_3$ quantum well structures. Section IV summarizes the design of SWIR and LWIR QWIPs and discusses estimated performance metrics. Section V summarizes the theoretical estimates of this work and offers future directions in light of these findings.

## II. Methods and Models

Band diagram and electron wavefunctions are calculated using the AQUILA MATLAB toolbox, developed by M. Rother [49]. AQUILA utilizes a user-defined material database, providing self-consistent solution of Schrödinger and Poisson equations for a user-defined structure under an applied electric field. The wavefunction and eigen energy solutions are compared to another Schrödinger-Poisson solver, Bandeng [50] and found to be consistent. β-$(Al_xGa_{1-x})_2O_3$ band gap energy and conduction band offsets as a function of Aluminum composition $x$ are found using Vegard's law, with bowing parameters determined by the theoretical work of T. Wang *et al* [33]. We assume a single band, isotropic electron effective mass that is linearly interpolated also using Vegard's law. The material properties used in AQUILA are summarized in Table I.

| Parameter | Value | Reference |
|---|---|---|
| $E_{g,Ga_2O_3}$ (eV) | 4.69 | [33] |
| $E_{g,Al_2O_3}$ (eV) | 7.03 | [33] |
| $E_{g,(Al_xGa_{1-x})_2O_3}$ (eV) | $4.69 + 1.34x + 1.0x^2$ | [33] |
| $\Delta E_c$ (eV) | $\begin{cases} 1.95x^2 - 0.24x + 0.96, & x > 0.5 \\ 0.94x^2 + 2.15x, & x < 0.5 \end{cases}$ | [33] |
| $m^*_{Ga_2O_3}$ (kg) | $0.28 m_0$ | [40] |
| $m^*_{(Al_xGa_{1-x})_2O_3}$ (kg) | $(0.28 + 0.11x) m_0$ | [40] |
| $\varepsilon_{r,Ga_2O_3}$ | 10 | [51] |
| $\varepsilon_{r,(Al_xGa_{1-x})_2O_3}$ | 10 | Assumed |

*Table 1: Material properties relevant to β-($Al_xGa_{1-x}$)$_2$O$_3$/Ga$_2$O$_3$ quantum wells used in this work.*

All simulation structures consist of a single Ga$_2$O$_3$ quantum well surrounded by two (Al$_x$Ga$_{1-x}$)$_2$O$_3$ barrier regions; total structure width is held constant at 200 Å. The range of peak transition wavelengths are calculated by varying the quantum well width $L_z$ from 2 to 14 nm for a given Aluminum composition $x$. For each width, $x$ is varied from 0.1 to 0.8. As noted previously in Section I, the conduction band discontinuity $\Delta E_c$ has only been experimentally verified for compositions up to $x = 0.2$ [35, 36]. At high compositions approaching $x = 0.7$ and above, it is possible that strain-induced effects could alter energy band diagrams considered in this work. The well region is undoped in solving for confined electron wavefunctions and eigen energies.

For this work, we only consider transitions from the ground state $|1\rangle$ to first excited state $|2\rangle$, for which the energy separation $E_{21} = E_2 - E_1$ determines the peak transition wavelength $\lambda_p = hc/E_{21}$. To calculate the absorption coefficient [25], we use Fermi Golden Rule. The oscillator strength between $|1\rangle$ and $|2\rangle$ states is represented by the dipole matrix element $M_{21}$, which is calculated from the wavefunction overlap integral [25]:

$$M_{21} = \langle 2|e \times z|1 \rangle = e \int_{-\infty}^{\infty} \psi_2^* \times z \times \psi_1 dz \quad (1)$$

where $e$ is the fundamental charge of the electron, $\psi_1$ and $\psi_2$ are the normalized wavefunctions of the first and second subbands, respectively, and $z$ is the direction perpendicular to device layers. In terms of $M_{21}$, the absorption coefficient is then expressed as [25]:

$$\alpha = \frac{2\pi c}{\lambda L_z} \sqrt{\frac{\mu_0}{\varepsilon_0 \varepsilon_r}} \sin^2\theta \, |M_{21}|^2 (n_1 - n_2) \times \frac{\hbar/\tau}{\left(\left(E_{21} - hc/\lambda\right)^2 + \left(\hbar/\tau\right)^2\right)} \quad (2)$$

where $c$ is the speed of light in vacuum, $\lambda$ is the wavelength of incident light, $\mu_0$ and $\varepsilon_0$ are the permeability and permittivity of free space, respectively, and $\varepsilon_r$ is high frequency dielectric constant of the well region. The $sin^2\theta$ term accounts for the polarization selection rule of intersubband transitions, namely only light polarized with an electric field component perpendicular to the quantum well plane is absorbed [52]. We assume $\theta = 45°$. The right-hand term is a Lorentzian broadening factor, with half width at half maximum (HWHM) equal to $\hbar/\tau$, where $\hbar$ is reduced Planck's constant and $\tau$ is the coherence decay time. The value of $\tau$ is influenced by several mechanisms including interface roughness scattering and optical phonon emission [53]. Values for $\tau$ are experimentally determined through pump and probe measurements or fitting of measured absorption spectra. Theoretically estimating $\tau$ for $Ga_2O_3$ quantum wells is beyond the scope of this work, so it is necessary to pick a representative range of $\tau$ (50 fs–5 ps) to comfortably represent unknowns in regard to the particular relaxation mechanisms of $\beta$-$Ga_2O_3$ quantum wells. The 2D carrier density of the $i$-th subband, $n_i$, is calculated from the 2D density of states and Fermi-Dirac statistics:

$$n_i = \frac{m^* kT}{\pi \hbar^2} \ln\left(\frac{1}{1-f_i}\right) \quad (3)$$

where $k$ is Boltzmann's constant, $T$ is the temperature, and $f_i$ is the occupation probability of the $i$-th subband. The fermi level is determined by the background-limited infrared performance (BLIP) condition $E_F = E_1 + kT$. The absorption probability $\eta$ is then calculated by considering the path length of light crossing a single quantum well region:

$$\eta = \frac{\alpha L_z}{\cos(\theta)} \quad (4)$$

The dark current is the current flowing through a biased detector, with no incident photon flux. For a given quantum well structure, the dark current versus applied field response is calculated using the Levine model [12]. The Levine model starts by determining an effective 3D density of electrons above the barrier from the thermal distribution of 2D electrons, multiplied by the tunneling transmission probability $T(E,F)$ for an electron at energy $E$ and applied electric field $F$ [52]:

$$n(F) = \left(\frac{m^*}{\pi \hbar^2 L_p}\right) \int_{E_1}^{\infty} f(E) T(E,F) dE \quad (5)$$

$L_p$ is the period width of one well and one barrier region. $f$ is the Fermi-Dirac distribution with fermi level $E_f = E_1 + kT$ given by the background-limited infrared performance (BLIP) condition. The transmission probability $T$ is calculated using WKB approximation [52]:

$$T(E,F) = \exp\left(-2 \int_0^{z_c} \sqrt{\frac{2m_b^*(V_0 - eFz - E)}{\hbar^2}} dz\right) \quad (6)$$

where $z_c$ is the classical turning point, $m_b^*$ is the electron effective mass of the barrier, $V_0$ is the potential barrier height of the well, and $E$ is the electron energy. Both $V_0$ and $E$ are referenced to the bottom of the well. The use of Eq. (6) only applies in the case of sufficiently slowly-varying potential, where WKB theory holds. This is valid for QWIP devices as the applied electric field is relatively low. The dark current density $J_{dark}$ is then calculated using the drift current equation:

$$J_{dark} = en(F)v(F) \quad (7)$$

The drift velocity $v(F)$ is calculated assuming a mobility $\mu$ that is interpolated across a range of recent experimental measurements of MOCVD-grown epitaxial thin films at varying temperatures [41]. We assume the mobility of the β-$(Al_xGa_{1-x})_2O_3$ barrier is the same as that of $Ga_2O_3$. Saturation velocity $v_{sat}$ is taken to be $10^7$ cm/s agreeing with theory and observed values [54, 55]:

$$v(F) = \frac{\mu F}{\sqrt{1 + \left(\frac{\mu}{v_{sat}}\right)^2}} \tag{8}$$

The photoconductive gain $g_{photo}$ of a QWIP describes the ratio of photoexcited electrons arriving at the collector contact to the number of absorbed photons. To calculate $g_{photo}$, it is necessary to consider the relevant processes for electron escape and capture from the well. A photoexcited electron in the $E_2$ state can escape from the well region via tunneling to the barrier. Thus, the escape time $\tau_{esc}$ is found using the transmission probability of Eq. (5), at an "attempt frequency" $v/2L_z$ [52]:

$$\tau_{esc} = \left(2\frac{L_z}{v}\right) T(E_2)^{-1} \tag{9}$$

The photoexcited electron may also relax back into the $E_1$ state before escaping the well. To describe this process, we take the intersubband relaxation time $\tau_{relax}$ to be the decoherence time used to describe absorption linewidth in Eq. (2).

A continuum electron travels above the barrier and crosses the quantum well period $L_p$ during some transit time $\tau_{trans}$, calculated using drift velocity of Eq. (6):

$$\tau_{trans} = \frac{L_p}{v(F)} \tag{10}$$

The continuum electron may not make it across the well region but instead scatter to the $E_1$ state i.e. capture into the well. The capture time $\tau_c$ is calculated as the continuum-to-bound scattering time via optical phonon emission. For the extended states near the top of the barrier, $\tau_c$ is

expressed as [52, 56]:

$$\tau_c = \frac{4h\Delta E_{co}L_p}{(g(E_{LO})+1)e^2 E_{LO}I_1}\left(\frac{1}{\epsilon_\infty}-\frac{1}{\epsilon_s}\right)^{-1} \tag{11}$$

where $h$ is Planck's constant, $\Delta E_{co}$ is the cutoff energy (defined as the energy difference between the top of the barrier and the $E_1$ state), $E_{LO}$ is the longitudinal optical (LO) phonon energy, $g(E_{LO})$ is the phonon occupation probability using Bose-Einstein statistics, $I_1$ is a dimensionless integral, and $\epsilon_\infty$ and $\epsilon_s$ are the high-frequency and low-frequency dielectric constants, respectively. The value for $E_{LO}$ is taken here to be 21 meV, which has proven to be a dominant mode in β-Ga$_2$O$_3$ at low electric fields [57]. The $I_1$ integral is approximately equal to 2 for the continuum-to-ground interaction, irrespective of the exact shape of the quantum well [56]. Eq. (11) assumes unconfined phonon modes and no screening of the electron-phonon interaction.

The probability that a photoexcited electron escapes the well and contributes to photocurrent is the escape probability $p_{esc}$ which is defined in terms of $\tau_{relax}$ and $\tau_{esc}$. Similarly, the capture probability is expressed by $\tau_{transit}$ and $\tau_c$ [52]:

$$p_{esc} = \frac{\tau_{relax}}{\tau_{relax}+\tau_{esc}} \tag{12}$$

$$p_c = \frac{\tau_{trans}}{\tau_c+\tau_{trans}} \tag{13}$$

$g_{photo}$ is defined using the ratio of $p_{esc}$ to $p_c$ [52]:

$$g_{photo} \equiv \frac{p_{esc}}{Np_c} \tag{14}$$

For these calculations, the number of quantum well periods $N$ is always 1. The responsivity $R$ describes the device current at a given incident light power. With an expression for $g_{photo}$, $R$ is calculated using the value of $\eta$ found in Eq. (4) [52]:

$$R = e\frac{\lambda}{hc}\eta g_{photo} \tag{15}$$

The specific detectivity $D^*$ offers a meaningful figure of merit that is comparable across QWIP designs. $D^*$ is defined as signal-to-noise ratio per incident power, normalized to detector area $A$ and measurement bandwidth $\Delta f$ [52]:

$$D^* = \frac{R\sqrt{A\Delta f}}{i_n} \quad (16)$$

The noise current $i_n$ of a QWIP receives dominant contributions from both thermal fluctuations of $J_{dark}$ as well as background photon noise. To calculate the background photon flux $\phi_{B,ph}$, we assume an ideal blackbody source [52]:

$$\phi_{B,ph} = \int \pi \sin^2 \frac{\theta}{2} \eta(\lambda) L_B \, d\lambda \quad (17)$$

A field of view full cone angle $\theta$ of 30° is used in this work, the photon irradiance $L_B$ is given by Planck's law at an assumed background temperature $T$ [52]:

$$L_B = \frac{2c}{\lambda^4} \frac{1}{e^{hc/\lambda kT} - 1} \quad (18)$$

It is important to note the difference in gain mechanisms between background noise and dark current noise, which is described in detail by Liu *et al* [58]. An electron excited into the excited |2⟩ state by a background photon has a probability $p_e$ of contributing to the noise current and thus is sufficiently described by $g_{photo}$. The escape mechanism of dark current is inherent within the Levine model as Eqn. (5) explicitly includes transmission probability to continuum, $T(E,F)$. Thus, the dark current noise gain $g_{noise}$ is solely expressed using $p_c$ [58]:

$$g_{noise} = \frac{1}{Np_c} \quad (19)$$

Where again, $N = 1$ for this single quantum well consideration. The dark current noise $i_{n,dark}$, background noise current $i_{n,b}$, and the expression for $D^*$ follows:

$$i_{n,dark}^2 = 4eg_{noise}J_{dark}A\Delta f \quad (20)$$
$$i_{n,B}^2 = 4e^2 g_{photo}^2 \phi_{B,ph} A\Delta f \quad (21)$$

$$D^* = \frac{R}{\sqrt{4eg_{noise}J_{dark}+4e^2 g_{photo}^2 \phi_{B,ph}}} \tag{22}$$

Note that Eq. (20) is proportional to $g_{noise}$ while Eq. (21) is proportional to the square of $g_{photo}$. This difference is based on the different gain mechanisms and is discussed in more detail by [58].

### III. The $(Al_xGa_{1-x})_2O_3/Ga_2O_3$ Design Space

As discussed in section I, the β-$(Al_xGa_{1-x})_2O_3/Ga_2O_3$ (AGO/GO) material system enables a large range of conduction band offsets (CBOs) and an expectedly large design space. The design space is found by plotting the peak absorption wavelength $\lambda_p$ (the wavelength at which Eq. (2) is maximized) vs $x$ for a family of well widths $L_z$. The β-AGO/GO design space for $x$ ranging from 0.05 to 0.7 and $L_z$ ranging from 2 nm to 14 nm is shown in Fig. 1. For a given barrier composition $x$, $\lambda_p$ increases with $L_z$ due to decreasing $E_{21}$. Each $L_z$ curve approaches the infinite-well energy separations as $x$ increases. At low $x$, the first excited state approaches the top of the barrier. Only bound states are considered in Fig. 1, which is the reason for the truncated $L_z = 2\ nm$ curve. When $x$ is further reduced, the $|2\rangle$ state becomes unbound.

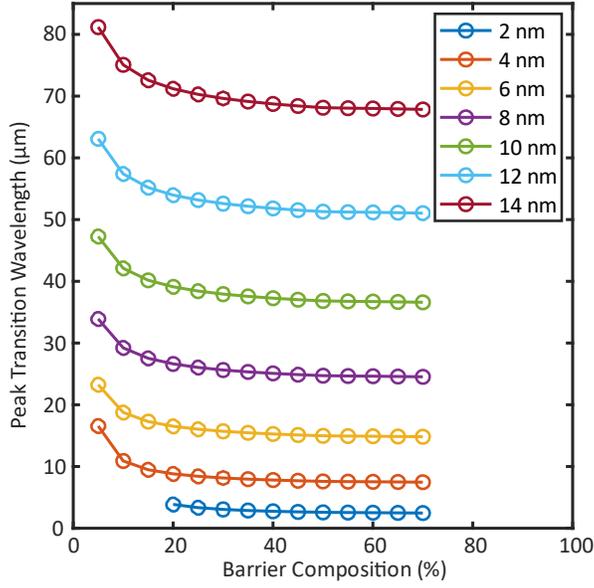

*Figure 1. Peak transition wavelength $\lambda_p$ versus Al composition in the barrier. The large CBO of AGO/GO heterojunction enables transition wavelengths as short as $\lambda_p$ = 2.5 μm for 70% barriers and as long as $\lambda_p$ = 81 μm for 5% barriers.*

The minimum detection wavelength is set by the maximum CBO of the material system. Since *β*-phase is predicted to be thermodynamically stable to 70% $Al_2O_3$ composition, this sets a maximum CBO of 1.7 eV. Fig. 2. compares the absorption spectra of a $(Al_{0.7}Ga_{0.3})_2O_3/Ga_2O_3$ quantum well structure with well widths again ranging from $L_z$ = 2 nm to $L_z$ = 14 nm. Large absorption coefficients on the order of $10^5$ cm$^{-1}$ are attained and peak absorption wavelengths spanning the entire infrared spectrum from $\lambda_p$ = 2.5 μm (short-infrared) up to $\lambda_p$ = 81 μm (far-infrared) are achieved. This result demonstrates the possibility of using a single, β-AGO/GO material system to design intersubband optoelectronic and electronic devices for a range of applications from night vision imaging arrays in the near-infrared to astronomical observation technology in the far-infrared. It should be noted that thermionic emission over the shallow barriers associated with long-wavelength devices tend to limit the performance of QWIPs. Thus, although Fig. 1. demonstrates transition wavelengths reaching 81 μm, devices designed in this regime would face the same issue of large dark currents as in conventional material systems.

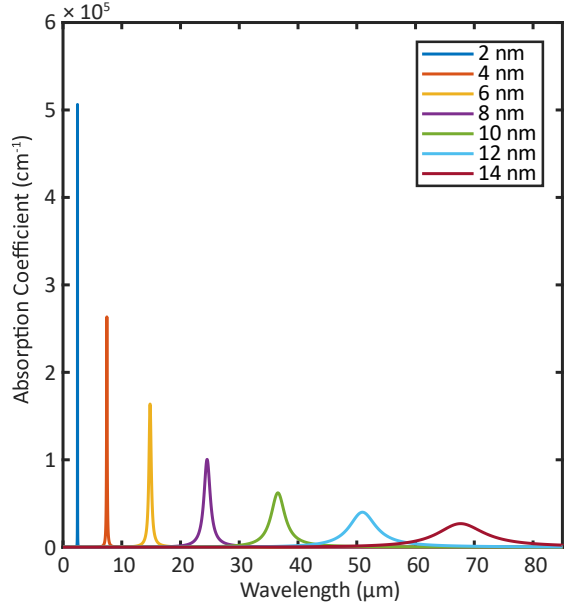

*Figure 2. Calculated absorption spectra for a single quantum well of varying width $L_z$ with $(Al_{0.7}Ga_{0.3})_2O_3$ barriers. Larger $L_z$ decreases $E_{21}$, enabling longer peak absorption wavelengths.*

## IV. Design of SWIR and LWIR QWIPs

To estimate QWIP performance throughout the design space, two $\beta\text{-}(Al_xGa_{1-x})_2O_3/Ga_2O_3$ QWIPs operating in the technologically significant short-wavelength infrared (SWIR) and long-wavelength infrared (LWIR) regimes are simulated. The SWIR QWIP is designed at the optical communication wavelength of $\lambda_p = 1.5$ *µ*m. The LWIR QWIP is designed for $\lambda_p = 9$ *µ*m, often used for thermal imaging applications. Each detector is designed with a bound-to-bound transition between two bound states. To enable efficient escape and collection at moderate electric fields, the barrier concentration is chosen such that the excited state |2⟩ is of reasonable energy separation from the top of the well (14.5 meV for SWIR detector and 12.8 meV for LWIR detector). The SWIR QWIP dimensions are chosen to be $L_z = 1.2$ nm and $x = 0.42$. The LWIR QWIP dimensions are of $L_z = 3.1$ nm and $x = 0.09$. Equilibrium band diagrams and electron wavefunctions, plotted at their corresponding eigen energies, are shown in Fig. 3. The dashed line represents the Fermi Level as dictated by the BLIP condition, described earlier in Section II.

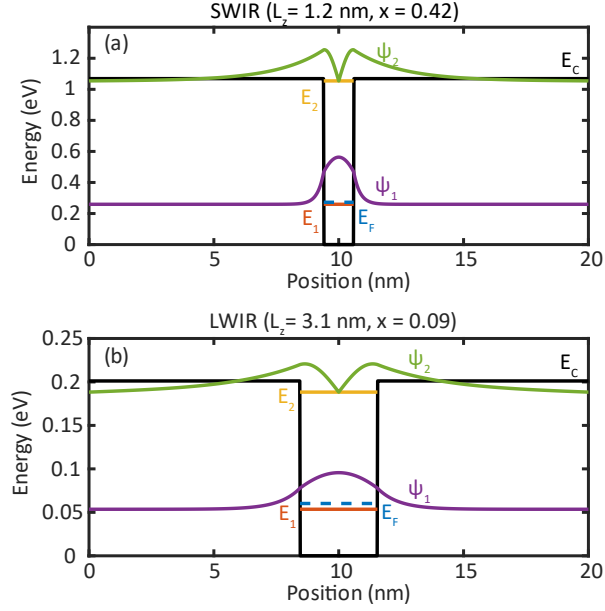

*Figure 3. Equilibrium band diagrams for (a) LWIR and (b) SWIR QWIPs. Electron wave functions are plotted along with their respective eigen energies. The dashed line represents the Fermi level $E_F$ as defined in the text.*

As evident from Eq. (1), $M_{21}$ is directly related to the wavefunction overlap integral. As wavefunction overlap decreases, the interaction weakens, and $M_{21}$ decreases proportionally. The effect of applied electric field on the dipole matrix element for the LWIR structure is shown in Fig. 4; the top of the barrier and $\psi_2$ wavefunction are plotted as an inset at electric field values $F$ = 10, 20, and 40 kV cm$^{-1}$. $\psi_2$ extends into the barrier regions at high fields, resulting in a reduction of $|M_{21}|$ with applied bias.

The effect of applied electric field on the dipole matrix element for the SWIR structure is shown in Fig. 5. The value of $|M_{21}|$ is about 30% lower and a downward slope is also seen at high fields. The decreased $|M_{21}|$ is explained by stronger confinement of the ground state wavefunction in the deep well of the SWIR design. Penetration depth $\delta_i$ of the $i$-th wavefunction into the classically forbidden region decreases as the energy separation from the top of the barrier, $U_0 - E_i$, increases [59]:

$$\delta_i = \frac{\hbar}{\sqrt{2m_e(U_0 - E_i)}} \tag{23}$$

Since $U_0 - E_2$ is similar between the SWIR and LWIR designs, the penetration depth of the excited state, $\delta_2$, differs only slightly between the two detectors. On the other hand, the potential well is much deeper for the SWIR device ($U_0 - E_1$ is about 809 meV compared to 148 meV in the LWIR design). The ground-state wave function is almost 3x as spread out in the LWIR device, with $\delta_1 =$ 9.5 Å compared to only about 3.8 Å in the SWIR detector. The value of $M_{21}$ for the SWIR detector will be smaller than the LWIR detector because the wave function overlap is in general less.

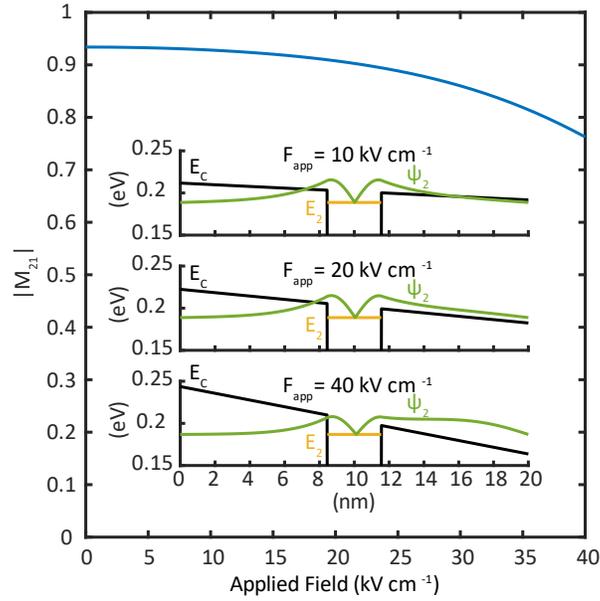

*Figure 4: (LWIR) Normalized dipole matrix element $|M_{21}|$ versus applied electric field. Inset figure shows the top of the quantum well and its excited state under three biasing conditions. The wave function spreads into the barrier region at high fields, decreasing wave function overlap and, consequently, $M_{21}$.*

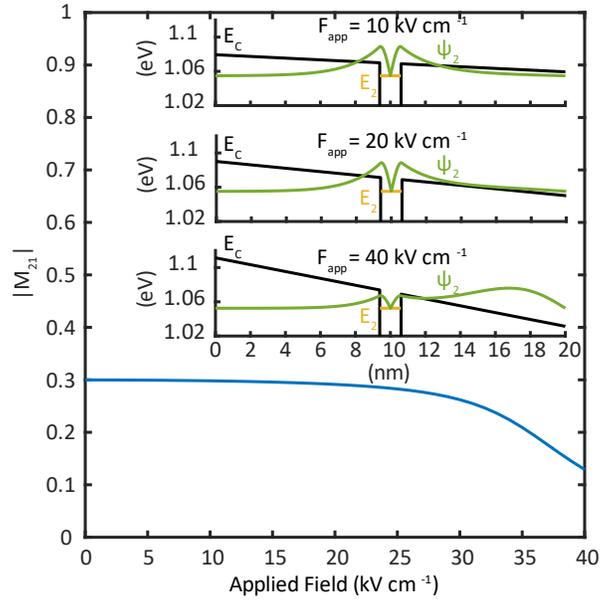

*Figure 5: (SWIR) Normalized dipole matrix element |M$_{21}$| versus applied electric field. Inset figure shows the top of the quantum well and its excited state under three biasing conditions. The wave function spreads into the barrier region at high fields, decreasing wave function overlap and, consequently, M$_{21}$.*

Dark current versus applied electric field is shown in Figs. 6 and 7 for the LWIR and SWIR detectors, respectively. Fig. 6 demonstrates the dramatic sensitivity of $J_{dark}$ on temperature which increases nearly 5 orders of magnitude from 0.003 A cm$^{-2}$ at $T = 77$ K to 112 A/cm$^2$ at $T = 150$ K ($F_{app} = 20$ kV cm$^{-1}$). When $T = 150$ K, $E_f$ is a value of $kT = 13$ meV above $E_1$ (BLIP condition, see section II). For the small CBO of the LWIR detector, this puts the top of the barrier at $E_f + 11kT$, and the Fermi-Dirac distribution of carriers contributing to dark current through Eq. (5) is appreciable. The large dark currents for $T \gg 77$ K in the LWIR design is the reason conventional QWIP detectors must be cryogenically cooled, otherwise the signal-to-noise ratio becomes unacceptable.

The SWIR detector also demonstrates an exponential increase in dark current, over 13 orders of magnitude, but at $T = 150$ K $J_{dark}$ is only $6.3 \times 10^{-21}$ A cm$^{-2}$ ($F_{app} = 20$ kV cm$^{-1}$)—nothing compared to that in the LWIR design. The reason behind highly suppressed $J_{dark}$ is the large CBO for $x = 0.42$ (1.07 eV) versus $x = 0.09$ (148 meV). The top of the barrier in the SWIR design is located at $E_f + 62kT$ and the number of thermally-excited carriers is comparably negligible. Thus, the SWIR design exhibits extremely low dark current values, $1.4 \times 10^{-7}$ A cm$^{-2}$, even at $T = 300$ K, which suggests background-limited performance may be achieved at room temperature, implying the possibility of uncooled short-wavelength QWIPs. It should be noted that the model used in Eqn. 7 assumes dark current is dominated by thermionic emission and thermionic field emission, which is typically the case for LWIR QWIPs given their small barrier heights. However, such models tend to underestimate dark current in LWIR QWIPs at very low temperatures where other tunneling mechanisms become prevalent [60]. In this work, we have not considered dark current contributions from defect-assisted tunneling, which have been experimentally demonstrated and theoretically investigated by others [61-64]. In reality, the calculated dark currents of Fig. 7 could be significantly higher due to trap-assisted tunneling via barrier region defects.

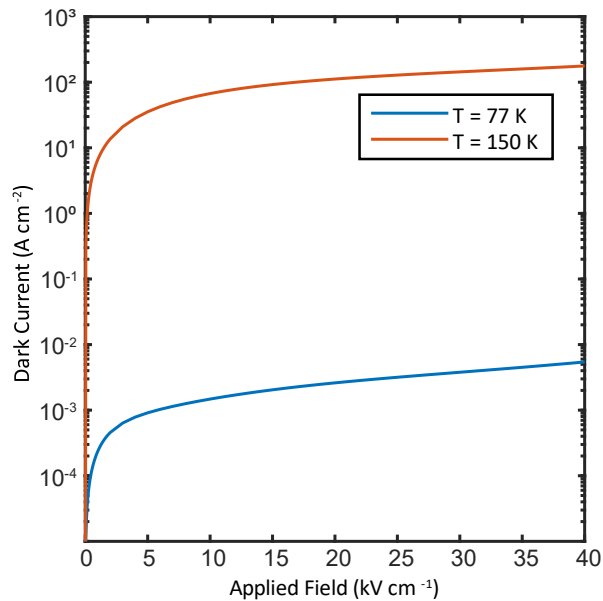

*Figure 6: (LWIR) Dark current density as a function of applied electric field, for T = 77 K and T = 150 K. Large dark currents at T > 77 K are the reason conventional LWIR QWIPs must be cryogenically cooled.*

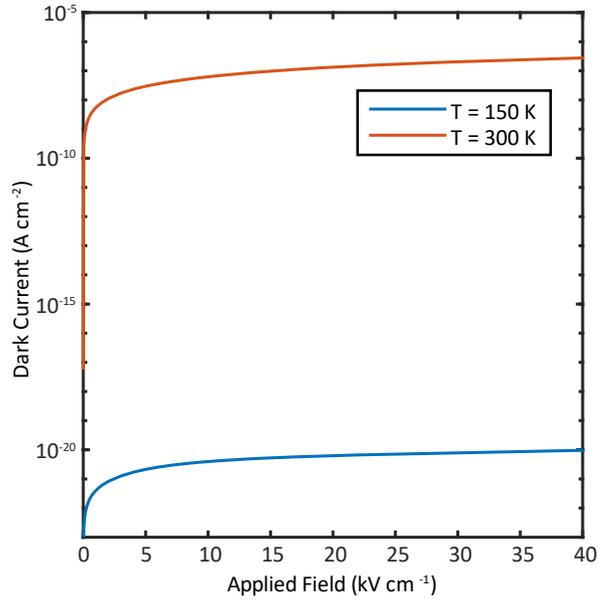

*Figure 7: (SWIR) Dark current density as a function of applied electric field, for T = 150 K and T = 300 K. Thermal contributions to dark current are dramatically suppressed due to large barrier height in the SWIR design.*

Background-limited performance is desired to achieve the best possible signal-to-noise ratio, and thus detectivity, in the detector. Background-limited operation is achieved when the noise contributions from dark current are equal to or less than that from background radiation i.e. when Eqs. (20) and (21) are equal [52]. In Fig. 8, peak responsivity and noise current contributions for the LWIR detector are plotted versus applied electric field at $T = 150$ K. The $R$ vs $F$ curve is typical for bound-to-bound devices—$R$ exhibits a flat region at low fields where the escape probability, $p_{esc}$, is zero, but increases once $p_{esc}$ becomes appreciable. As $g_{photo}$ increases, $R$ follows this trend but peaks as $\psi_2$ spreads into the barrier region at high $F$. Decreasing dipole strength causes $R$ to drop at high $F$. At T = 150 K, $i_{n,dark}$ is the dominant contributor to noise, more than 4 orders of magnitude larger than $i_{n,b}$. This is not surprising considering the large dark currents previously noted in Fig. 6. At $T = 150$ K, the LWIR QWIP is dark-current limited.

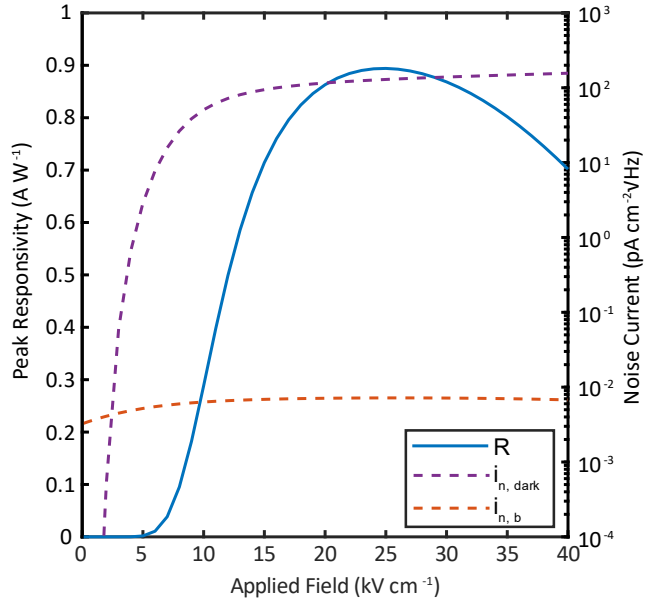

*Figure 8: (LWIR) On the left axis (solid line), peak responsivity versus applied electric field. On the right axis (dashed lines), noise current contributions from both dark current and background flux. Operating temperature is 150 K. The LWIR detector is dark-current limited.*

In Fig. 9, peak responsivity and noise current contributions for the SWIR detector are plotted versus applied electric field at $T = 150$ K. Peak responsivity is lower because of the smaller $M_{21}$, as discussed previously. Since dark current is greatly suppressed, the dominant noise contribution is switched for the SWIR detector. It is observed that $i_{n,b}$ is about 3 orders of magnitude greater than $i_{n,dark}$. At $T = 150$ K the SWIR QWIP is background-limited.

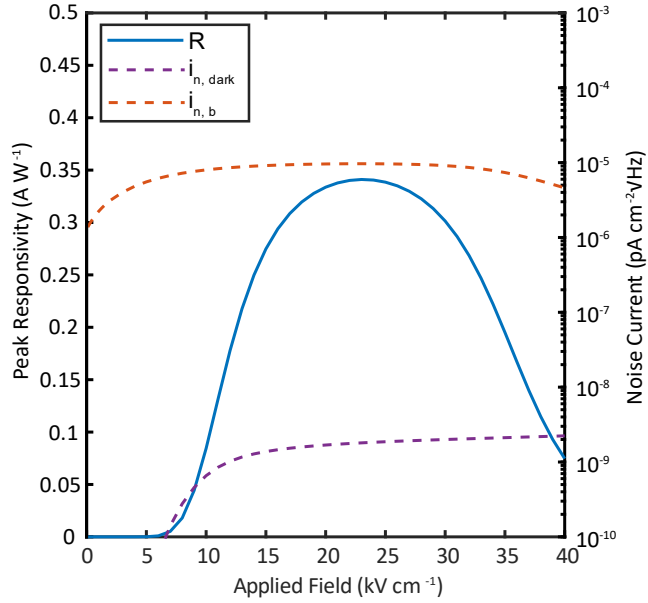

*Figure 9: (SWIR) On the left axis (solid line), peak responsivity versus applied electric field. On the right axis (dashed lines), noise current contributions from both dark current and background flux. Operating temperature is 150 K. The SWIR detector is background-limited.*

Peak detectivity $D^*$ versus applied field is plotted in Fig. 10 for the LWIR QWIP at $T = 77$ K and $T = 150$ K. $D^*$, like responsivity, also reaches a peak value. Peak $D^*$ is $8.64 \times 10^9$ cm Hz$^{1/2}$/W at $T = 77$ K and $7.61 \times 10^7$ cm Hz$^{1/2}$/W at $T = 150$ K. Detectivities of conventional QWIPs are around $10^9 - 10^{11}$ cm Hz$^{1/2}$/W; it is clear that the LWIR design must be cryogenically cooled to achieve conventional performance.

With a larger barrier to thermally-excited electrons, the SWIR QWIP is expected to offer higher detectivities. The peak $D^*$ vs $F$ curve for the SWIR detector is shown in Fig. 11 at $T = 150$ K and $T = 300$ K. Peak $D^*$ is $3.52 \times 10^{14}$ cm Hz$^{1/2}$/W at $T = 150$ K, which is nearly 7 orders of magnitude larger than the LWIR detector at the same temperature. Such large detectivities still hold at room temperature, with $D^*$ reaching $5.84 \times 10^{11}$ cm Hz$^{1/2}$/W at $T = 300$ K. As noted before, $J_{dark}$, and thus $i_{n,dark}$ could be much greater in considering additional dark current mechanisms. Even if $J_{dark} = 10^{-10}$ A cm$^{-2}$, peak $D^*$ only reduces to $1.6 \times 10^{13}$ cm Hz$^{1/2}$/W at $T = 150$ K and $5.8 \times 10^{11}$ cm Hz$^{1/2}$/W at $T = 300$ K.

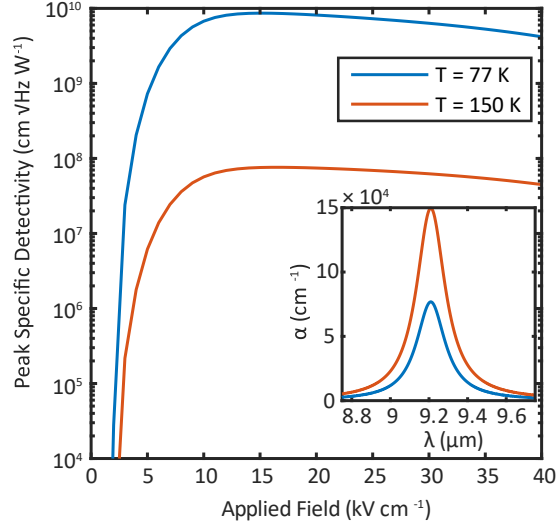

*Figure 10: (LWIR) Peak specific detectivity as a function of applied electric field for T = 77 K and T = 150 K. Inset figure shows the corresponding absorption spectra. $D^*_{peak} = 8.64 \times 10^9$ cm $Hz^{1/2}$/W at 77 K and $7.61 \times 10^7$ cm $Hz^{1/2}$/W at 150 K.*

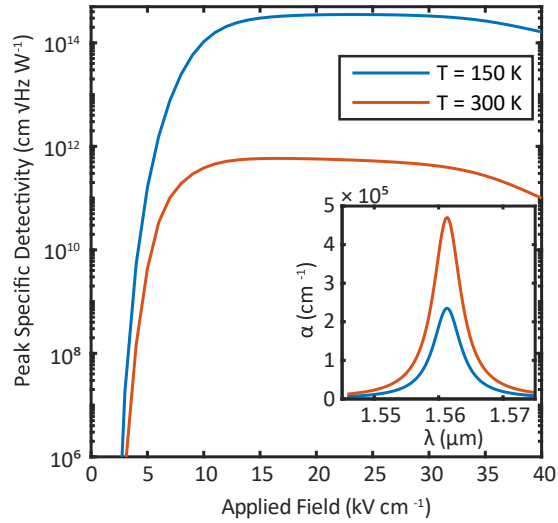

*Figure 11: (SWIR) Peak specific detectivity as a function of applied electric field for T = 150 K and T = 300 K. Inset figure shows the corresponding absorption spectra. $D^*_{peak} = 3.52 \times 10^{14}$ cm $Hz^{1/2}$/W at 150 K and $5.84 \times 10^{11}$ cm $Hz^{1/2}$/W at 300 K.*

Peak detectivity versus temperature is compared between measurements of existing LWIR and SWIR QWIPs in Figs. 12 and 13. The error bars express the range of detectivities for the range of $\tau$ which was assumed in Eqn. 2 (the lower error bar corresponds to $\tau = 50$ fs and the upper error bar to τ = 5 ps). Fig. 12 includes some conventional materials for LWIR [13, 65-68] as well as a novel approaches using the wide bandgap II-VI ZnCdSe alloy system [20]. Fig. 13 includes attempts at achieving short-wavelength QWIPs by using wide bandgap materials [19, 26], GaInAsN alloys [69], and the use of strained double barrier structures [70]. It is important to note that the referenced detectors utilize multiple quantum well designs each with a different number of quantum wells *N*. Figs. 12 and 13 compare single quantum well detectivities $D^{*\,(1)}$ where detectivity values of each design have been normalized to *N* [52]:

$$D^{*(1)} \cong \frac{D^{*(N)}}{\sqrt{N}} \tag{24}$$

Fig. 12 confirms comparable cryogenic performance to existing LWIR QWIPs while Fig. 13 predicts superior SWIR performance of AGO/GO, demonstrating the potential of high quality ultrawide bandgap heterostructures for SWIR QWIPs.

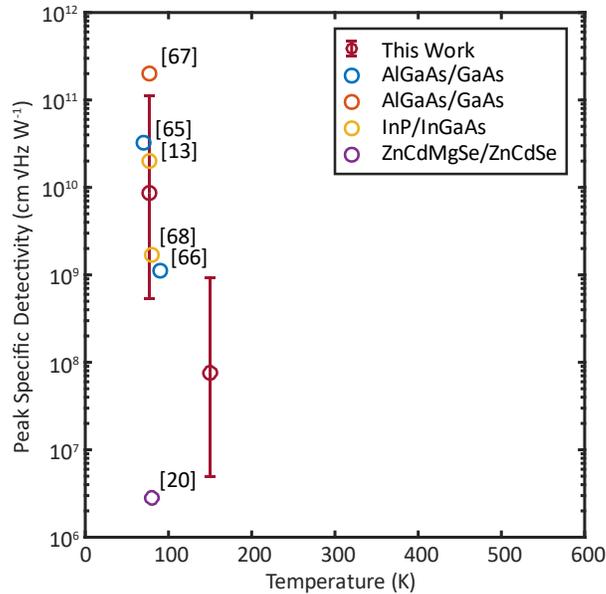

*Figure 12: (LWIR) Maximum specific detectivity versus operating temperature, plotted against the same metric of existing LWIR QWIPs. The lower error bar represents the peak detectivity when τ = 50 fs and the top error bar represents τ = 5 ps.*

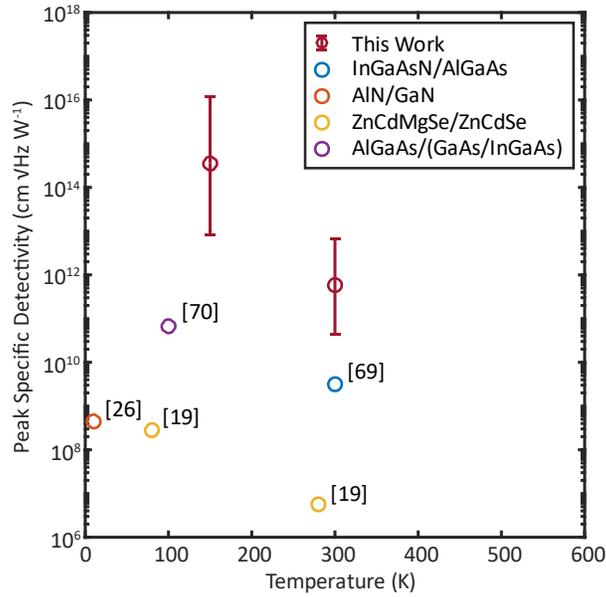

*Figure 13: (SWIR) Maximum specific detectivity versus operating temperature, plotted against the same metric of existing SWIR QWIPs. The lower error bar represents the peak detectivity when τ = 50 fs and the top error bar represents τ = 5 ps.*

This work demonstrates the promise of Gallium Oxide-based intersubband devices. As a first-principles approach, there are several areas needing further development. Most prominently, a stricter consideration of scattering mechanisms in Gallium Oxide materials is needed to fully understand decoherence and relaxation processes. Thorough theoretical and experimental investigation of dark current mechanisms in $Ga_2O_3$ materials is also necessary. With promising results on a single quantum well QWIP, the opportunity to explore optimized doping, operating temperature, and MQW designs is a future pursuit. The first experimental confirmation of infrared absorption in AGO/GO quantum wells is necessary to confirm these theoretical findings. Beyond the QWIP, an expansive design space of heterostructures implies opportunities for other intersubband devices, namely RTDs, QCLs, and QCDs which remain to be explored.

## V.     Conclusion

This work constitutes a theoretical consideration of intersubband transitions designed in ultra-wide bandgap $Ga_2O_3$ quantum wells. An expansive β-$(Al_xGa_{1-x})_2O_3$/$Ga_2O_3$ design space ranging from short-wavelength ($\lambda_p = 2.5$ $\mu$m) to far ($\lambda_p = 81$ $\mu$m) infrared regions is demonstrated through numerical simulation and first-principles modeling. The performance characteristics of two technologically relevant QWIPs are also estimated. A LWIR ($\lambda_p \sim 9.2$ $\mu$m) QWIP shows comparable performance to conventional material systems, while a SWIR ($\lambda_p \sim 1.6$ $\mu$m) QWIP offered higher detectivities than previously realized in other materials. The demonstration of room temperature detectivity on the order of $10^{11}$ cm Hz$^{1/2}$/W at wavelengths as short as the optical communication wavelength offers extremely exciting prospects for QWIP applications. The implication of uncooled QWIP detectors is immensely exciting for novel device applications where bulky refrigeration equipment associated with conventional QWIPs is unfeasible such as integrated detectors for optical communication systems or compact imaging systems.

## VI.     Acknowledgements

This material is based upon work supported by the Air Force Office of Scientific Research under award number FA9550-18-1-0507 monitored by Dr Ali Sayir. Any opinions, findings and conclusions or recommendations expressed in this material are those of the authors and do not necessarily reflect the views of the United States Air Force. This work was supported by funding from the Undergraduate Research Opportunities Program at the University of Utah awarded to Joseph E. Lyman.